\documentclass[twoside,english]{paper}
\usepackage[T1]{fontenc}
\usepackage[latin9]{inputenc}
\usepackage{geometry}
\usepackage{babel}
\usepackage{mathrsfs}
\usepackage{amsmath}
\usepackage{amsthm}
\usepackage{amssymb}
\usepackage{booktabs}
\usepackage{array}
\usepackage{rotating}
\usepackage{caption}

\usepackage[pdfusetitle,
 bookmarks=true,bookmarksnumbered=false,bookmarksopen=false,
 breaklinks=false,pdfborder={0 0 1},backref=false,colorlinks=false]
 {hyperref}

\makeatletter
\let\paperclassexample\example

\let\example\relax
\AtBeginDocument{%
  \@ifundefined{example}{\let\example\paperclassexample}{}
}
\numberwithin{equation}{section}
\numberwithin{figure}{section}
\newcommand{\lyxaddress}[1]{
	\par {\raggedright #1
	\vspace{1.4em}
	\noindent\par}
}

\makeatother

\begin{document}
\title{The Dirac equation in (split-)octonions: origins, variants,
and modern context}
\subtitle{20 October 2026 (v18)}
\author{Jens K\"oplinger}
\maketitle

\lyxaddress{jenskoeplinger@gmail.com}
\begin{abstract}
Octonions and split-octonions have been used to express the Dirac
equation in physics in several conceptually distinct ways. This review
organizes them into four groups: the \emph{2-factor}, \emph{3-factor},
and \emph{projection} representations, which use (split-)octonion basis
elements natively to model a spacetime basis, and the \emph{conventional}
use of octonions to carry Dirac algebra acting on \emph{spinors}. For each
group we identify the originating construction, relate later variants to
it, and point to modern applications and to recent methods for native
split-octonionic analysis. Because the multiplication table of the
(split-)octonions is not unique, forms that look different in print can
coincide after a norm-preserving linear isometry and a
multiplication-preserving change of basis; we make such relations
explicit and tabulate the basis conventions used across select sources.
The 2-factor representation is treated in most detail: we document its
origin and show that a recently proposed split-octonionic Dirac equation
coincides with it after such an isometry and change of basis, while
crediting the independent contributions of that work.
\end{abstract}
\tableofcontents{}

\pagebreak{}

\section{Split-octonions (hyperbolic octonions)}

Split-octonions $\mathbb{O}^{\prime}$ are eight-dimensional, nonassociative
unital composition algebras \cite{Jacob1958} that are typically understood
over the real numbers. Any number $x\in\mathbb{O}^{\prime}$ is uniquely
represented as a vector $x=\left(x_{0},\ldots,x_{7}\right)\in\mathbb{R}^{8}$
after fixing a vector space basis of $\mathbb{R}^{8}$. Subsequently,
the algebra $\left\langle \mathbb{O}^{\prime}\mid*\mid+\right\rangle $
is specified by requiring existence of a multiplicative unit $e_{0}\in\mathbb{O}^{\prime}$
and fixing a multiplication rule that satisfies product-composition
of a quadratic form $\left\Vert \cdot\right\Vert $:
\begin{align}
\left\Vert \cdot\right\Vert  & :\mathbb{O}^{\prime}\rightarrow\mathbb{R},\\
\left\Vert x\right\Vert  & :=\left(x_{0}^{2}+x_{1}^{2}+x_{2}^{2}+x_{3}^{2}\right)-\left(x_{4}^{2}+x_{5}^{2}+x_{6}^{2}+x_{7}^{2}\right),\label{eq:defQuadraticForm}\\
\left\Vert x*y\right\Vert  & =\left\Vert x\right\Vert \left\Vert y\right\Vert \textrm{ for any }x,y\in\mathbb{O}^{\prime}.
\end{align}
Here, the vector space basis is chosen such that $x_{0}$ through
$x_{3}$ contribute with a positive sign to the quadratic form, and
$x_{4}$ through $x_{7}$ contribute with a negative sign. By convention,
the $0$-dimension is typically identified with the real number line,
such that $e_{0}:=\left(1,0,\ldots,0\right)$ is the multiplicative
unit. These choices are far from unique, and many equivalent
representations for split-octonion multiplication exist that satisfy
these constraints. We will not go into detail around possible choices,
algebra isomorphisms and automorphisms (``symmetries'') here. For
reviews see e.g.~\cite{Baez2002,CS2003,Okubo1995}.

Looking at the definition of the quadratic form (\ref{eq:defQuadraticForm}),
the fact that the number of positive and negative signs split the
eight coefficients in two groups of four gives rise to the commonly
used modern terminology ``split-octonions''. The same indefinite form
$\left\Vert \cdot\right\Vert $ also contains hyperbolic planes in various
two-dimensional subspaces, hence the equivalent terminology ``hyperbolic
octonions'' for the same algebra.

\section{The Dirac equation in (split-)octonions: four approaches}

The Dirac equation in physics is the fundamental equation of motion
of a free spin-$\frac{1}{2}$ particle. It uses the particle's mass
$m\in\mathbb{R}$, partial differential operators $\partial_{\mu}:=\partial/\partial x_{\mu}$
(with $\mu=0,1,2,3$) in space $x_{1},x_{2},x_{3}\in\mathbb{R}$ and
time $x_{0}\in\mathbb{R}$, and a complex-valued wave function that
is a function of spacetime, $\Psi:\mathbb{R}^{4}\rightarrow\mathbb{C}^{4}\cong\mathbb{R}^{8}$,
$\Psi=\left(\psi_{0}^{\mathrm{r}}+i\psi_{0}^{\mathrm{i}},\ldots,\psi_{3}^{\mathrm{r}}+i\psi_{3}^{\mathrm{i}}\right)$,
where $\psi_{\mu}^{\mathrm{r}}$ denotes the real part of the $\mu$-component
of $\Psi$ and $\psi_{\mu}^{\mathrm{i}}$ its imaginary part. Customarily
written by means of Dirac matrices $\gamma^{\mu}$, which satisfy
$\gamma^{\mu}\gamma^{\nu}+\gamma^{\nu}\gamma^{\mu}=2\eta^{\mu\nu}$ with
$\eta^{\mu\nu}=\mathrm{diag}(1,-1,-1,-1)$ and thus generate a matrix
representation of the corresponding Clifford algebra, the equation becomes
\begin{equation}
\left(\sum_{\mu=0}^{3}i\gamma^{\mu}\partial_{\mu}-m\right)\Psi=0.\label{eq:DiracEqn}
\end{equation}
Throughout we adopt natural units $\hbar=c=1$; conventions around
metric signature, index notation, summation, and the metric tensor
$\eta_{\mu\nu}$ vary across the literature, and
we give only a simplified overview, referring to the canonical literature
for details. This linear equation can be separated into eight real
expressions.

In a broad context, there are four conceptually different ways in
which (split-)octonions have been used to express the Dirac equation.
The \emph{2-factor}, \emph{3-factor}, and \emph{projection} representations
use (split-)octonion basis elements natively to model a spacetime
basis, which introduces nonassociativity in a way that requires clarification
when connecting to the canonical formulation of physical law. In contrast,
the \emph{conventional} approach uses octonions to carry ordinary
Dirac algebra acting on \emph{spinors}.\footnote{Given almost 100 years
of knowledge of both the Dirac equation and split-octonions, it is
well possible that further relevant work is not properly acknowledged
here. Despite our best efforts, we sincerely apologize for any oversight
and ask to be notified and given the chance to make amends.}

\subsection{The 2-factor representation}

The \emph{2-factor representation} writes the eight real expressions of
the Dirac equation as a single product of two hypercomplex
numbers,
\begin{equation}
\nabla\Psi=0,
\end{equation}
the wave function $\Psi$ being one (split-)octonion and the first-order
operator $\nabla$ another, carrying the mass, with all components
pairwise identified.

The earliest attempt, by R.~Penney \cite{Penney1968}, used the
\emph{real} (division) octonions $\mathbb{O}$. He took the wave function
to be a single octonion $\psi$ and wrote the left product
\begin{equation}
\Big(e_{0}\partial_{0}+\sum_{k=1}^{3}e_{k}\partial_{k}+\sum_{n=4}^{7}e_{n}m_{n}\Big)\psi=0,
\end{equation}
with $e_{0}=1$ and four real mass parameters, constrained by
$m^{2}=m_{4}^{2}+m_{5}^{2}+m_{6}^{2}+m_{7}^{2}$, in place of the single
Dirac mass. Every imaginary unit squares to $-1$, generating a Euclidean
Clifford structure, and not a Minkowski one. The eight real equations
do not close to one Dirac equation, but to a \emph{coupled pair}
with extra mass-like terms. Penney returned to the problem in a later
paper \cite{Penney1971}, whose equation is covariant and follows from a
Lagrangian, and which gains an isospin doublet as an internal degree of
freedom, from right-multiplication of the wave function by the Pauli
matrices.

A \emph{complex}-octonion case was proposed by F.~Colombo,
I.~Sabadini and D.~C.~Struppa \cite{CSS2000}, in the 16-dimensional
$\mathbb{C}\otimes\mathbb{O}$: they take $\psi$ a single complex octonion
and write
\begin{equation}
\Big(iA_{0}\partial_{0}+i\sum_{\nu=1}^{3}A_{\nu}\partial_{\nu}-\sum_{\mu=4}^{7}A_{\mu}m_{\mu}\Big)\psi=0,
\end{equation}
each $A_{i}$ being the $8\times8$ matrix of left-multiplication by the
$i$-th element of their basis $a_{0}=e_{0}$, $a_{r}=ie_{r}$
($r=1,\ldots,7$), whose non-real elements square to $+1$; the matrix
representation of such one-sided octonion multiplication is described
in detail by G.~M.~Dixon \cite{Dixon1994}.
These matrices, collected into the operator $\mathcal{D}$, are the
regular representation of that single left
product, making it a 2-factor form,
$\mathcal{D}\psi=0$ with $\mathcal{D},\psi\in\mathbb{C}\otimes\mathbb{O}$.
The imaginary $i$ now supplies the signature the real octonions lacked,
and a Klein--Gordon equation follows. The mass again is allowed to split
into four real components $m_{4},\ldots,m_{7}$.

As an algebra, the \emph{split} octonions $\mathbb{O}^{\prime}$ supply
both metric signs from single units: after choosing a basis
$b_{\mathbb{O}^{\prime}}:=\left\{ 1,i_{1},i_{2},i_{3},\varepsilon_{4},\varepsilon_{5},\varepsilon_{6},\varepsilon_{7}\right\} $,
the $i_{1},i_{2},i_{3}$ can be taken to square to $-1$ and the
$\varepsilon_{4},\ldots,\varepsilon_{7}$ to $+1$, matching the four positive
and four negative signs in the quadratic form (\ref{eq:defQuadraticForm}).
This is what lets the
full Dirac equation close as such a product, on the left alone. The
single mass enters as the scalar part of $\nabla$. This form was given
in 2006 by the present author \cite{Koepl2006}, with wave function and
operator
\begin{align}
\Psi & =\psi_{0}^{\mathrm{r}}+i_{1}\psi_{0}^{\mathrm{i}}+i_{2}\psi_{1}^{\mathrm{r}}+i_{3}\psi_{1}^{\mathrm{i}}+\varepsilon_{4}\psi_{2}^{\mathrm{r}}-\varepsilon_{5}\psi_{2}^{\mathrm{i}}-\varepsilon_{6}\psi_{3}^{\mathrm{r}}-\varepsilon_{7}\psi_{3}^{\mathrm{i}},\\
\nabla & =-m+i_{1}\partial_{0}-\varepsilon_{5}\partial_{3}+\varepsilon_{6}\partial_{2}-\varepsilon_{7}\partial_{1}\label{eq:nabla2006intro}
\end{align}
(detailed in Section~\ref{sec:detail}). In contrast to the two previous attempts
above, all spacetime derivatives are assigned to \emph{non-real} basis
units, which are pairwise anticommutative, now matching all required
properties of canonical Dirac algebra. To our knowledge, this
is the first description of the Dirac equation as a single split-octonion
product that meets them all. Enlarging
this
split-octonion construction to the sixteen-dimensional complex octonions
allows to introduce additional physical degrees of
freedom, namely a mixing angle, a four-potential, and an electric charge,
all in the 2-factor representation. This is
demonstrated in Section~\ref{sec:emergent}.

\subsection{The 3-factor representation\label{sec:threefactor}}

The \emph{3-factor representation} after S.~De~Leo and K.~Abdel-Khalek
\cite{DLAK1996a,DLAK1996b} writes the Dirac operator as a sum of
terms, each generally built from \emph{three} octonion basis
elements, in the real (division) octonions. In their form these terms
use two-sided multiplication, both left- and right-acting
(``barred'') octonionic operators, schematically
\begin{equation}
\Big(\sum_{k}a_{k}\,\square\,b_{k}\,\partial_{k}-m\Big)\Psi=0,
\end{equation}
where $\square$ marks the slot for the operand $\Psi$, acted on from
the left by $a_{k}$ and from the right by $b_{k}$. M.~Gogberashvili
\cite{Gogb2006a,Gogb2006b} built an analogous one-sided construction in
the split-octonions, acting on the wave function $\psi$ from the
\emph{left} alone:
\begin{equation}
\big[c\,\partial_{t}+J^{n}\partial_{n}\big]\psi+B\,(I\psi)=0,
\end{equation}
with $J_{n}$ the split-octonion units, $I$ the pseudoscalar, and
$B=-\left(mc+\tfrac{e}{c}A_{0}\right)+\tfrac{e}{c}A_{n}J^{n}$ the combined
mass and electromagnetic-potential term. The bar appears in the second
term: nonassociativity fixes the order, so that $I$ multiplies $\psi$
first. For nonvanishing vector potential, the term $\tfrac{e}{c}A_{n}J^{n}$
makes $B$ non-scalar, so that $B\,(I\psi)\neq(BI)\,\psi$ and the oriented
product does not reduce to left-multiplication by a single split octonion;
Gogberashvili notes it is ``similar to the barred operators''
of \cite{DLAK1996a,DLAK1996b}. In the free case, $B=-mc$ is a real scalar
and the product does collapse, to
$\big[c\,\partial_{t}+J^{n}\partial_{n}-mcI\big]\psi=0$. For $m\neq0$ this
does not share the standard Dirac mass-shell; its norm reduces to the ordinary
d'Alembertian only in the massless limit, as \cite{Gogb2006b} notes.
The line was later given a mathematical treatment by R.~Beradze
and T.~Shengelia \cite{BS2016}. De~Leo and Abdel-Khalek, moreover,
embed their construction in a broader program on a distinct nonassociative
quantum mechanics \cite{DLAK1996a}.

\subsection{The projection representation}

A different program has its roots in work by A.~Sudbery (with K.~W.~Chung)
\cite{CS1987,Sudb1984}, who use octonion basis elements in a $2\times2$
matrix setting in a treatment primarily in terms of Lie algebra. Starting
from a massless octonionic momentum-space equation in $10$ spacetime
dimensions \cite{SM1996}, T.~Dray and C.~A.~Manogue \cite{DM2000,MD1999} fix a
preferred octonionic unit
$\ell$ and project onto the complex subalgebra it generates,
\begin{equation}
\pi(q)=\tfrac{1}{2}\left(q+\ell\,q\,\bar{\ell}\right),
\end{equation}
reducing it to the (associative) Dirac equation (\ref{eq:DiracEqn}), with
the mass emerging from the projected momentum; the resulting spectrum is
interpreted as three particle families together with a sterile-neutrino-like
singlet.

\subsection{Conventional Dirac algebra on octonionic spinors}

In contrast to the three native representations above, octonions can
also be used to carry the \emph{conventional} Dirac algebra: the equation
keeps its standard form (\ref{eq:DiracEqn}),
with the $\gamma^{\mu}$ and the spinor $\Psi$ realized over an octonionic
structure, rather than the spacetime basis itself being modeled by
octonion units. This appeared as early as the papers of I.~Bengtsson
and M.~Cederwall \cite{BC1988} and R.~Foot and G.~C.~Joshi \cite{FJ1988},
in the context of supersymmetry.
S.~Marques-Bonham \cite{MarBon1991} develops the Dirac equation in this
conventional form over a split-octonion structure within a geometric
(non-Riemannian) field theory, and weaves it into a Lagrangian: the
octonionic Dirac equation follows from a covariant action principle whose
Lagrangian density is built from octonionic $\gamma$-matrices and
covariant derivatives, coupling the spin-$\tfrac{1}{2}$ field to
gravitational, electromagnetic, and octonionic Yang-Mills-like potentials
at once, with the standard Dirac equation in these fields recovered as a
limit. Works
by G.~M.~Dixon \cite{Dixon1994},
J.~Schray and C.~Manogue \cite{SM1996}, J.~C.~Baez and J.~Huerta
\cite{BH2010}, and N.~Furey \cite{Furey2012,Furey2016,Furey2025}
are just a few prominent modern examples of the program that seeks
Standard-Model structure from division-algebra representations, all of
which could benefit a concrete understanding of spacetime dynamics,
around the historical step from first- to second-quantization.

\subsection{Modern context and analytic methods}

Octonionic Dirac constructions are sometimes cited interchangeably despite
belonging to the distinct approaches above; two recent works illustrate this. A work by J.~C.~V\'elez Qui\~nones \cite{VQ2022}, ``in terms of $\mathrm{C}\ell_{2}$ Clifford algebra projectors,'' uses a
framework closer to Furey's minimal-left-ideal construction than to the
projection of \cite{DM2000,MD1999}, and cites \cite{DLAK1996b} incidentally. Relatedly, a work coauthored by Manogue and Dray
\cite{MDW2022} cites their projection approach \cite{DM2000,MD1999}
although that projection is not central there; their construction
of Lie algebras from one-sided octonionic multiplication may rather hint
at the 2-factor representation \cite{Koepl2006} when looking for
spacetime representations in not necessarily associative subspaces. Such
references to the Dirac equation
are made ``in passing'' and are not substantial for the respective
findings.

Finally, we point to recent mathematical methods for native split-octonionic
analysis \cite{HRS2025,KLL2025,KLL2025weyl,Lopat2026,LR2025,LZ2024},
which could make many of the above symbolic expressions on nonassociative
algebra analytically tractable. This includes a large body of contemporary
work \cite{CBN2011,CSN2015,CT=00005C"OD2008,Demir2013,DTK2013,Gogb2006a,KHG2025,Koepl2023a,Koepl2023b,Lasen2024,MBCM2020,TKD2012,TKD2014,Weng2009},
to mention just a few.

\section{The 2-factor representation in detail: origin and verification\label{sec:detail}}

We now treat the 2-factor representation in detail. It is the
representation in which the spacetime derivatives are carried by
pairwise-anticommuting non-real units $u_{\mu}$; in
(\ref{eq:nabla2006intro}) these are
$u_{0},\ldots,u_{3}=i_{1},-\varepsilon_{7},\varepsilon_{6},-\varepsilon_{5}$.
Since the split-octonions are alternative, the left-multiplication
operators $L_{\mu}x:=u_{\mu}x$ satisfy
$L_{\mu}L_{\nu}+L_{\nu}L_{\mu}=L_{u_{\mu}u_{\nu}+u_{\nu}u_{\mu}}=-2\eta_{\mu\nu}$,
the Clifford anticommutation relations of the coefficients
$i\gamma^{\mu}$ in (\ref{eq:DiracEqn}), and thus generate a real
representation on $\mathbb{R}^{8}$ of the Clifford algebra defined by
these relations.\footnote{In the interest of
transparency: this author published the 2006 form discussed here and is,
with M.~Gogberashvili, a coauthor of the 2008 essay
\cite{KDG2008} cited below. The equivalence established here is independently
checkable from the multiplication rules of \cite{GG2024} and \cite{Koepl2006}.}

\subsection{J.~K\"oplinger, \textquotedblleft Dirac equation on hyperbolic
octonions\textquotedblright{} (2006)}

Any split-octonion product can be separated into eight real
expressions, simply by writing out the component summations for each
of the eight dimensions. C.~Mus\`es had conjectured \cite{Muses1980}
that ``a simpler version of [Dirac's] equation using only 16-dimensional
$M$-algebra'' (his complex octonions) ``is possible''; no proof was given
there. In 2006 the paper \cite{Koepl2006} demonstrated
that it is possible to write the eight expressions of the Dirac equation
in symbolic form as a single split-octonion product, using
only two factors $\nabla\Psi=0$, by pairwise identifying all components.
We adopt throughout the corrigendum-corrected (2007) form of \cite{Koepl2006}.\footnote{The
correction follows \cite{Koepl2007c} footnote~2; for a self-contained
corrigendum with acknowledgment see e.g.~the ResearchGate brief \href{https://doi.org/10.13140/RG.2.2.22900.41602}{DOI:~10.13140/RG.2.2.22900.41602}
or the author's personal publications web page \cite{KoeplWWW}. The
original subscript ``$\mathrm{hyp8}$'' is omitted as it does not
add value here.} Written to split-octonion basis element $b_{\mathbb{O}^{\prime}}=\left\{ 1,i_{1},i_{2},i_{3},\varepsilon_{4},\varepsilon_{5},\varepsilon_{6},\varepsilon_{7}\right\} $,
equations (3), (4), and (5) in that publication are:
\begin{align}
\Psi & :=\left(\psi_{0}^{\mathrm{r}},\psi_{0}^{\mathrm{i}},\psi_{1}^{\mathrm{r}},\psi_{1}^{\mathrm{i}},\psi_{2}^{\mathrm{r}},-\psi_{2}^{\mathrm{i}},-\psi_{3}^{\mathrm{r}},-\psi_{3}^{\mathrm{i}}\right),\label{eq:koepl2006psi}\\
\nabla & :=\left(-m,\partial_{0},0,0,0,-\partial_{3},\partial_{2},-\partial_{1}\right),\label{eq:koepl2006DiracOp}\\
\nabla\Psi & =0.\label{eq:koepl2006DiracEqn}
\end{align}

\subsection{J.~K\"oplinger, V.~Dzhunushaliev, M.~Gogberashvili, \textquotedblleft Emergent
time from non-associative quantum theory\textquotedblright{} (2008)\label{sec:emergent}}

In an effort to connect the initial finding into existing related
context, the author collaborated on an essay contest
\cite{KDG2008}. The goal was to raise awareness of the research domain,
make it more accessible and venture a thought-provoking advertisement
for split-octonions in physics in general.

The essay bridged various referenced works. It also used the more
modern terminology ``split-octonion'' and clarified where alternate
terminology ``hyperbolic octonions'' arises from:
\begin{quotation}
``Just as the octonions, the split-octonions have a multiplicative
2-form; only this time it is not Euclidean, but hyperbolic, \ldots{}
The hyperbolic 2-form of split-octonions (20) \ldots ''
\end{quotation}
Equations (27), (28), and (29) therein refer to the follow-on
work \cite{Koepl2007c} that takes advantage of the simplicity of the
split-octonion product representation of the Dirac equation
\cite{Koepl2006}, to introduce expansions in the form of a mixing
angle $\alpha\in\mathbb{R}$, four-potential $\left(A_{0},\ldots,A_{3}\right)$
and charge $e$. Labeling octonion basis elements explicitly as $b_{\mathbb{O}}=\left\{ 1,i_{1},\ldots,i_{7}\right\} $
and a central $i_{0}$ with $i_{0}^{2}=-1$,
these equations are:
\begin{align}
\nabla_{\mathrm{A}} & :=i_{1}\left(\partial_{0}-i_{0}eA_{0}\right),\nonumber \\
\nabla_{\mathrm{B}} & :=i_{5}\left(\partial_{3}-i_{0}eA_{3}\right)+i_{6}\left(-\partial_{2}+i_{0}eA_{2}\right)+i_{7}\left(\partial_{1}-i_{0}eA_{1}\right),\\
\nabla & :=\nabla_{\mathrm{A}}+e^{i_{0}\alpha}\nabla_{\mathrm{B}},\nonumber
\end{align}
\begin{align}
\Psi_{\mathrm{A}} & :=\psi_{0}+i_{1}\psi_{1}+i_{2}\psi_{2}+i_{3}\psi_{3},\nonumber \\
\Psi_{\mathrm{B}} & :=-i_{4}\psi_{4}+i_{5}\psi_{5}+i_{6}\psi_{6}+i_{7}\psi_{7},\\
\Psi & :=\Psi_{\mathrm{A}}+e^{i_{0}\alpha}\Psi_{\mathrm{B}},\nonumber
\end{align}
\begin{equation}
\left(\nabla-m\right)\Psi=0.
\end{equation}
The reader easily verifies that this reduces to the 2006 representation
of the Dirac equation (\ref{eq:koepl2006DiracEqn}) when setting $\alpha=\frac{\pi}{2}$,
the potentials $\left(A_{0},\ldots,A_{3}\right)$ to zero, trivially
relabeling the real wave function components, and pairwise identifying
split-octonion basis elements $b_{\mathbb{O}^{\prime}}=\left\{ 1,i_{1},i_{2},i_{3},-i_{0}i_{4},\ldots-i_{0}i_{7}\right\} \equiv\left\{ 1,i_{1},i_{2},i_{3},\varepsilon_{4},\varepsilon_{5},\varepsilon_{6},\varepsilon_{7}\right\} $.
The essay cites this as ``{[}7{]}''.

\subsection{M.~Gogberashvili, A.~Gurchumelia, \textquotedblleft Split octonionic
Dirac equation\textquotedblright{} (2024)}

Split-octonions in \cite{GG2024} are written to basis elements
\begin{equation}
b_{\mathbb{O}^{\prime}}=\left\{ 1,j_{1},j_{2},j_{3},I,J_{1},J_{2},J_{3}\right\} .\label{eq:gogGurch2024basis}
\end{equation}
A wave function is initially defined in equation (4) as
\begin{equation}
\psi=\left(\begin{array}{c}
\psi_{4}+i\psi_{7}\\
-\psi_{6}+i\psi_{5}\\
\psi_{3}+i\psi_{0}\\
\psi_{1}+i\psi_{2}
\end{array}\right)
\end{equation}
and later (14) embedded into $\mathbb{O}^{\prime}$ as
\begin{equation}
\psi=\psi_{0}+I\psi_{4}+\sum_{n=1}^{3}\left(j_{n}\psi_{n}+J_{n}\psi_{4+n}\right).\label{eq:gogGurch2024psi}
\end{equation}
This associates one real component of the wave function with one split-octonion
basis, and the function is then assumed ``constant in variables $x_{0}$,
$x_{5}$, $x_{6}$ and $x_{7}$'', i.e., $\psi:\mathbb{R}^{4}\rightarrow\mathbb{R}^{8}$
just as in (\ref{eq:koepl2006psi}) above.

A differential operator $\mathscr{D}$ (15) acts on the variables
$x_{1}$, $x_{2}$, $x_{3}$ and $x_{4}\equiv t$,
\begin{equation}
\mathscr{D}=I\partial_{t}-\sum_{n=1}^{3}j_{n}\partial_{n},\label{eq:gogGurch2024DiracOp}
\end{equation}
such that the split-octonionic Dirac equation can be represented as
(16)
\begin{equation}
\left(\mathscr{D}-J_{3}m\right)\psi=0,\label{eq:gogGurch2024DiracEqn}
\end{equation}
which the authors then show to be ``component-wise equivalent to
the standard Dirac system'' with the help of a computer program.
The structure of (\ref{eq:gogGurch2024DiracEqn}) is that of a 2-factor
product of split-octonions. Left-multiplying the operator $\mathscr{D}-J_{3}m$
of (\ref{eq:gogGurch2024DiracEqn}) by $J_{3}$ and right-multiplying $\psi$
(\ref{eq:gogGurch2024psi}) by $J_{3}$, that is, the change of variables
$\mathscr{D}-J_{3}m\mapsto J_{3}\mathscr{D}-m$, $\psi\mapsto\psi J_{3}$, carries the
product $\left(\mathscr{D}-J_{3}m\right)\psi$ to its image under the map
$w\mapsto\left(J_{3}w\right)J_{3}$ of $\mathbb{O}^{\prime}$, a
norm-preserving linear isometry ($J_{3}^{2}=+1$, determinant $+1$). After a
signed, multiplication-preserving change of basis, the coordinates of the new
factors $J_{3}\mathscr{D}-m$ and $\psi J_{3}$ are, up to a renaming of the
eight real components, exactly those of the 2006 operator and wave function
(\ref{eq:koepl2006DiracOp}), (\ref{eq:koepl2006psi}).
The explicit derivation is given in Appendix~\ref{app:equiv}, which also
shows that the authors' Lagrangian (their equation~(17)) reduces to one
quarter of the standard Dirac Lagrangian.

Two observations follow. First, after this change of variables and change
of basis, (\ref{eq:gogGurch2024DiracEqn}) coincides term by term with the
2-factor equation of \cite{Koepl2006} in the same split-octonion
multiplication convention, so the authors' component-wise check amounts to an
independent verification. Second, their remark that the 2024 expressions ``are
slightly different from earlier findings'' in \cite{Gogb2006b}, with
``the main difference [being] the appearance of split octonionic imaginary
unit $J_{3}$,'' is accounted for by the same change of variables: the
appearance of $J_{3}$ is absorbed by it, and is not a substantive difference
between them. The
genuine structural difference, which the authors do not emphasize, is with the
earlier construction \cite{Gogb2006b}, distinct from both: it is the one-sided,
oriented (nonassociatively bracketed) left-action operator analogous to the
De~Leo--Abdel-Khalek barred operators
\cite{DLAK1996a,DLAK1996b} (Section~\ref{sec:threefactor}), a two-term
form that carries the time derivative on the real unit and the mass on
the pseudoscalar $I$, whereas (\ref{eq:gogGurch2024DiracEqn}) is a
2-factor product with all derivatives on non-real, pairwise-anticommuting
units.

Beyond the equivalence, \cite{GG2024} contributes in its own right: an
independent component-wise verification carried out with the open-source
\texttt{SplitOct} package, and an embedding of the split-octonionic Dirac
equation into the authors' broader Lagrangian program \cite{GG2023}.

\section*{Acknowledgments}

I am grateful
for all the impulses, pointers, inspiration, help, support, and references
in the early days that led to the original 2006 paper. It is an honor
to have written and published in the context of those who helped me,
to have acknowledged them in the relevant publications and talks over
the years \cite{KoeplWWW}, and to now place that work into a wider
context in the form of a review. I am also thankful for a large amount
of personal feedback received since the early versions of this review,
for detailed anonymous reviewer feedback that substantially strengthened
the mathematics, and to several mentors for their
guidance on how to protect an academic claim while at the same time
improving on the literature.

\appendix

\section{Explicit reduction of the 2024 Dirac equation and Lagrangian\label{app:equiv}}

We demonstrate explicitly that the split-octonionic Dirac equation
(\ref{eq:gogGurch2024DiracEqn}) is equivalent to the 2-factor representation
(\ref{eq:koepl2006DiracEqn}) through an invertible norm-preserving linear
transformation together with a multiplication-preserving signed change of
basis. We form a new operator $\mathscr{D}^{\prime}$ by left-multiplying
$\mathscr{D}$ (\ref{eq:gogGurch2024DiracOp}) with $J_{3}$, using the
multiplication rules stated in \cite{GG2024}:
\begin{equation}
\mathscr{D}^{\prime}:=J_{3}\mathscr{D}=j_{3}\partial_{t}+J_{2}\partial_{1}-J_{1}\partial_{2}-I\partial_{3}.\label{eq:appDprime}
\end{equation}
Right-multiplying $\psi$ (\ref{eq:gogGurch2024psi}) with $J_{3}$ yields:
\begin{align}
\psi^{\prime} & :=\psi J_{3}\label{eq:appPsiprime}\\
 & =J_{3}\psi_{0}-j_{3}\psi_{4}+J_{2}\psi_{1}-j_{2}\psi_{5}-J_{1}\psi_{2}+j_{1}\psi_{6}-I\psi_{3}+\psi_{7}.
\end{align}
Because split-octonions satisfy the Moufang property
$\left(zx\right)\left(yz\right)=\left(z\left(xy\right)\right)z$, and
$J_{3}^{2}=+1$ (so $J_{3}^{-1}=J_{3}$, with $\left\Vert J_{3}\right\Vert =-1$),
we obtain
\begin{equation}
\left(\mathscr{D}^{\prime}-m\right)\psi^{\prime}=\left(J_{3}\mathscr{D}-J_{3}\left(J_{3}m\right)\right)\left(\psi J_{3}\right)=\left(J_{3}\left(\left(\mathscr{D}-J_{3}m\right)\psi\right)\right)J_{3}.\label{eq:appMoufang}
\end{equation}
Thus $\left(\mathscr{D}-J_{3}m\right)\psi=0$ if and only if
$\left(\mathscr{D}^{\prime}-m\right)\psi^{\prime}=0$, because the map
$T\left(w\right):=\left(J_{3}w\right)J_{3}$ is an invertible norm-preserving
linear isometry of $\mathbb{O}^{\prime}$ with determinant
$+1$.\footnote{For norm-preserving transformations of split-octonion space
in general see e.g.~\cite{Gogb2008}; the automorphism group of
$\mathbb{O}^{\prime}$, the non-compact form of $G_{2}$, to which $T$ does not
belong, is studied by the authors of \cite{GG2024} themselves in \cite{GG2019}.
The map $T$ is not an automorphism of $\mathbb{O}^{\prime}$: for example,
$T\left(j_{1}\right)=-j_{1}$ and $T\left(j_{2}\right)=-j_{2}$, so
$T\left(j_{1}\right)T\left(j_{2}\right)=j_{1}j_{2}=j_{3}$ in the convention
of \cite{GG2024}, whereas $T\left(j_{1}j_{2}\right)=T\left(j_{3}\right)=-j_{3}$.
The identity (\ref{eq:appMoufang}) is an instance of the Moufang property:
the two factors $\mathscr{D}-J_{3}m$ and $\psi$ are transformed separately,
by left and by right multiplication with $J_{3}$, and the product of the
transformed factors $J_{3}\left(\mathscr{D}-J_{3}m\right)=\mathscr{D}^{\prime}-m$
and $\psi J_{3}=\psi^{\prime}$ equals the transform $T$ of the original
product $\left(\mathscr{D}-J_{3}m\right)\psi$.}

To compare the transformed equation with
(\ref{eq:koepl2006psi})--(\ref{eq:koepl2006DiracEqn}), we use the signed
reordered basis
$\tilde{b}_{\mathbb{O}^{\prime}}:=\left\{ 1,j_{3},-j_{2},j_{1},-J_{3},I,-J_{1},-J_{2}\right\} $
(Table~\ref{tab:splitoct-crossref} lists this basis, and those of the
other works discussed, coordinate by coordinate). The multiplication rules
of \cite{GG2024}, compared with the corrigendum-corrected table of
\cite{Koepl2006}, show that the correspondence
\begin{equation}
\left\{ 1,i_{1},i_{2},i_{3},\varepsilon_{4},\varepsilon_{5},\varepsilon_{6},\varepsilon_{7}\right\} \mapsto\left\{ 1,j_{3},-j_{2},j_{1},-J_{3},I,-J_{1},-J_{2}\right\} \label{eq:appCorr}
\end{equation}
preserves multiplication; for example,
\begin{equation}
j_{3}\left(-j_{2}\right)=j_{1},\quad j_{3}\left(-J_{3}\right)=I,\quad\left(-j_{2}\right)\left(-J_{3}\right)=-J_{1},\quad j_{1}\left(-J_{3}\right)=-J_{2},\label{eq:appExamples}
\end{equation}
and the first three non-real elements square to $-1$ while the last four
square to $+1$. In this basis, the coordinate vectors of $\psi^{\prime}$ and
$\mathscr{D}^{\prime}-m$ are
\begin{align}
\psi^{\prime} & =\left(\psi_{7},-\psi_{4},\psi_{5},\psi_{6},-\psi_{0},-\psi_{3},\psi_{2},-\psi_{1}\right),\\
\mathscr{D}^{\prime}-m & =\left(-m,\partial_{t},0,0,0,-\partial_{3},\partial_{2},-\partial_{1}\right).
\end{align}
After renaming the eight real component functions, and with
$\partial_{t}=\partial_{0}$, these are exactly the coordinate vectors $\Psi$
and $\nabla$ in (\ref{eq:koepl2006psi}) and (\ref{eq:koepl2006DiracOp}).
Hence the transformed equation
$\left(\mathscr{D}^{\prime}-m\right)\psi^{\prime}=0$ is the 2006 2-factor
equation (\ref{eq:koepl2006DiracEqn}), term by term, in the same
split-octonion multiplication convention.

The same invertible change of variables, (\ref{eq:appDprime})--(\ref{eq:appPsiprime}),
carries forward to the Lagrangian.
The authors \cite{GG2024} further give a split-octonionic Lagrangian,
their equation~(17),
\begin{equation}
\mathcal{L}=\tfrac{1}{2}\left\langle J_{3}\psi,\mathscr{D}\psi\right\rangle +\tfrac{1}{2}m\left\langle \psi,\psi\right\rangle ,
\end{equation}
formed with the bilinear form $\left\langle \cdot,\cdot\right\rangle $ that
polarizes the split-octonion norm (their equation~(9)). In the variables of
(\ref{eq:appDprime})--(\ref{eq:appPsiprime}) it reads
\begin{equation}
\mathcal{L}=\tfrac{1}{2}\left\langle \psi^{\prime},\left(\mathscr{D}^{\prime}-m\right)\psi^{\prime}\right\rangle ,\label{eq:appLagTransformed}
\end{equation}
since $\langle J_{3}\psi,\mathscr{D}\psi\rangle =\langle \psi^{\prime},\mathscr{D}^{\prime}\psi^{\prime}\rangle $
and $\langle \psi,\psi\rangle =-\langle \psi^{\prime},\psi^{\prime}\rangle $.
Under the identification of $\psi$ with the standard
Dirac spinor $\Psi$ of (\ref{eq:DiracEqn}), the split-octonion
norm is exactly the Lorentz scalar, $\left\langle \psi,\psi\right\rangle =-\overline{\Psi}\Psi$; and the factor $J_{3}$
supplies the Dirac adjoint $\overline{\Psi}=\Psi^{\dagger}\gamma^{0}$, so that
$\left\langle J_{3}\psi,\mathscr{D}\psi\right\rangle $ is the standard Dirac kinetic
term. Equation~(17) then equals
\begin{equation}
\mathcal{L}=\tfrac{1}{4}\left[\overline{\Psi}\left(i\gamma^{\mu}\partial_{\mu}-m\right)\Psi+\mathrm{h.c.}\right],
\end{equation}
one quarter of the standard Dirac Lagrangian. The
Lagrangian is therefore the standard Dirac object expressed in split-octonions,
as its corresponding equation of motion (\ref{eq:gogGurch2024DiracEqn}) is; the construction does not leave $\mathbb{O}^{\prime}$,
and the unit $J_{3}$ is again absorbed by the change of variables.

\begin{table}[p]
\centering
\rotatebox{90}{%
\begin{minipage}{\textheight}
\footnotesize
\setlength{\tabcolsep}{3pt}
\renewcommand{\arraystretch}{1.3}
\captionof{table}{Cross-reference of split-octonion basis-element notation and
coordinate alignment across the publications discussed, ordered by year. The eight
numbered columns are the coordinates $0,\ldots,7$ of $\mathbb{R}^{8}$; each cell
gives, in the notation of the work in its row, the signed basis element placed at
that coordinate. The rows are signed bases aligned for comparison in the present
manuscript, and not necessarily in the native order in which the cited works
introduce their basis elements; each row, read as a correspondence with the
reference row, preserves the multiplication table.\label{tab:splitoct-crossref}}
\par\smallskip
\centering
\begin{tabular}{@{}m{34mm} *{8}{>{\centering\arraybackslash}m{8mm}} c m{58mm}@{}}
\toprule
 & & \multicolumn{3}{c}{imaginary} & \multicolumn{4}{c}{split} & & \\
\cmidrule(lr){3-5}\cmidrule(lr){6-9}
Publication & $0$ & $1$ & $2$ & $3$ & $4$ & $5$ & $6$ & $7$ & Ref & Notes \\
\midrule
K\"oplinger 2006
 & $1$ & $i_1$ & $i_2$ & $i_3$ & $\varepsilon_4$ & $\varepsilon_5$ & $\varepsilon_6$ & $\varepsilon_7$
 & \cite{Koepl2006}
 & Reference basis and order. \\
\addlinespace[3pt]
Gogberashvili 2006
 & $1$ & $j_3$ & $-j_2$ & $j_1$ & $-J_3$ & $I$ & $-J_1$ & $-J_2$
 & \cite{Gogb2006b}
 & Signed alignment to \cite{Koepl2006}. \\
\addlinespace[3pt]
K\"oplinger--Dzhunushaliev--Gogberashvili 2008
 & $1$ & $i_1$ & $i_2$ & $i_3$ & $-i_0i_4$ & $-i_0i_5$ & $-i_0i_6$ & $-i_0i_7$
 & \cite{KDG2008}
 & Split units $\varepsilon_n{=}{-}i_0i_n$. \\
\addlinespace[3pt]
Gogberashvili 2009
 & $1$ & $j_3$ & $-j_2$ & $j_1$ & $-J_3$ & $I$ & $-J_1$ & $-J_2$
 & \cite{Gogb2008}
 & Same signed alignment. \\
\addlinespace[3pt]
Gogberashvili--Gurchumelia 2024
 & $1$ & $j_3$ & $-j_2$ & $j_1$ & $-J_3$ & $I$ & $-J_1$ & $-J_2$
 & \cite{GG2024}
 & Signed basis $\tilde{b}_{\mathbb{O}^{\prime}}$ used in (\ref{eq:appCorr}). \\
\bottomrule
\end{tabular}
\end{minipage}%
}
\end{table}


\clearpage

\begin{thebibliography}{KLL2025weyl}

\bibitem[Baez2002]{Baez2002}J.~C.~Baez, The Octonions, \emph{Bull.~Amer.~Math.~Soc.~}\textbf{39}
(2002), 145-205; \href{https://arxiv.org/abs/math/0105155}{\emph{arXiv}:math/0105155v4~[math.RA]}.

\bibitem[BC1988]{BC1988}I.~Bengtsson, M.~Cederwall, Particles,
twistors and the division algebras, \emph{Nucl.~Phys.~B} \textbf{302}
(1988), 81-103.

\bibitem[BH2010]{BH2010}J.~C.~Baez, J.~Huerta, ``Division Algebras
and Supersymmetry I'' in \emph{Superstrings, Geometry, Topology,
and C{*}-algebras}, eds.~R.~Doran, G.~Friedman and J.~Rosenberg;
\emph{Proc.~Symp.~Pure Math.}~\textbf{81} (2010) 65-80; \href{https://arxiv.org/abs/0909.0551}{\emph{arXiv}:0909.0551~[hep-th]}.

\bibitem[BS2016]{BS2016}R.~Beradze, T.~Shengelia, Dirac and Maxwell
equations in Split Octonions, \emph{arXiv} (2016); \href{https://arxiv.org/abs/1610.06418}{\emph{arXiv}:1610.06418~[physics.gen-ph]}.

\bibitem[CBN2011]{CBN2011}B.~C.~Chanyal, P.~S.~Bisht, O.~P.~S.~Negi,
Generalized Split-Octonion Electrodynamics, \emph{Int.~J.~Theor.~Phys.}\textbf{~50}
(2011), 1919--1926.

\bibitem[CSN2015]{CSN2015}B.~C.~Chanyal, V.~K.~Sharma, O.~P.~S.~Negi,
Octonionic Gravi-Electromagnetism and Dark Matter, \emph{Int.~J.~Theor.~Phys.}\textbf{~54}
(2015), 3516--3532.

\bibitem[CS1987]{CS1987}K.~W.~Chung, A.~Sudbery, Octonions and
the Lorentz and Conformal Groups of Ten-dimensional Space-time, \emph{Phys.~Lett.~B}\textbf{
198} (1987), 161-164.

\bibitem[CS2003]{CS2003}J.~H.~Conway, D.~Smith, On Quaternions
and Octonions (2003), A~K~Peters, Natick, MA.

\bibitem[CSS2000]{CSS2000}F.~Colombo, I.~Sabadini, D.~C.~Struppa,
Dirac Equation in the Octonionic Algebra, in \emph{Analysis, Geometry,
Number Theory: The Mathematics of Leon Ehrenpreis}, eds.~E.~L.~Grinberg,
S.~Berhanu, M.~Knopp, G.~Mendoza, E.~T.~Quinto; \emph{Contemporary
Math.}~\textbf{251} (2000), 117--134.

\bibitem[CT\"OD2008]{CT=00005C"OD2008}N.~Candemir, M.~Tan\i\c{s}l\i,
K.~\"Ozda\c{s}, S.~Demir, Hyperbolic Octonionic Proca-Maxwell
Equations, \emph{Z.~Naturforsch.}~\textbf{63a} (2008), 15-18.

\bibitem[Demir2013]{Demir2013}S.~Demir, Hyperbolic Octonion Formulation
of Gravitational Field Equations, \emph{Int.~J.~Theor.~Phys.} \textbf{52}
(2013), 105-116.

\bibitem[Dixon1994]{Dixon1994}G.~M.~Dixon, Division Algebras: Octonions,
Quaternions, Complex Numbers and the Algebraic Design of Physics (1994),
Kluwer Academic Publishers, Dordrecht.

\bibitem[DLAK1996a]{DLAK1996a}S.~De~Leo, K.~Abdel-Khalek, Octonionic
Quantum Mechanics and Complex Geometry, \emph{Prog.~Theor.~Phys.}\textbf{~96}
(1996) 823-831; \href{https://arxiv.org/abs/hep-th/9609032}{\emph{arXiv}:hep-th/9609032}.

\bibitem[DLAK1996b]{DLAK1996b}S.~De~Leo, K.~Abdel-Khalek, Octonionic
Dirac Equation, \emph{Prog.~Theor.~Phys.}\textbf{~96} (1996) 833--845;
\href{https://arxiv.org/abs/hep-th/9609033}{\emph{arXiv}:hep-th/9609033}.

\bibitem[DM2000]{DM2000}T.~Dray, C.~A.~Manogue, Quaternionic Spin,
in \emph{Clifford Algebras and their Applications in Mathematical
Physics}, eds. R.~Ab\l{}amowicz, B.~Fauser (Birkh\"auser, Boston,
2000); \href{https://arxiv.org/abs/hep-th/9910010}{\emph{arXiv}:hep-th/9910010}.

\bibitem[DTK2013]{DTK2013}S.~Demir, M.~Tan\i\c{s}l\i, M.~E.~Kansu,
Generalized Hyperbolic Octonion Formulation for the Fields of Massive
Dyons and Gravito-Dyons, \emph{Int.~J.~Theor.~Phys.}~\textbf{52}
(2013), 3696-3711.

\bibitem[FJ1988]{FJ1988}R.~Foot, G.~C.~Joshi, On a certain supersymmetric
identity and the division algebras, \emph{Mod.~Phys.~Lett.~A} \textbf{3}
(1988), 999-1004.

\bibitem[Furey2012]{Furey2012}C.~Furey, Unified Theory of Ideals,
\emph{Phys.~Rev.~D} \textbf{86} (2012), 025024; \href{https://arxiv.org/abs/1002.1497}{\emph{arXiv}:1002.1497~[hep-th]}.

\bibitem[Furey2016]{Furey2016}C.~Furey, ``Standard model physics
from an algebra?'', \emph{Ph.D.~Thesis} (2016); \href{https://arxiv.org/abs/1611.09182}{\emph{arXiv}:1611.09182~[hep-th]}.

\bibitem[Furey2025]{Furey2025}N.~Furey, A Superalgebra Within: Representations
of Lightest Standard Model Particles Form a $\mathbb{Z}_{2}^{5}$-Graded
Algebra, \emph{Ann.~Phys.}~(2025) e00229, \href{https://doi.org/10.1002/andp.202500229}{DOI:~10.1002/andp.202500229};
\href{https://arxiv.org/abs/2505.07923}{\emph{arXiv}:2505.07923~[hep-ph]}.

\bibitem[GG2019]{GG2019}M.~Gogberashvili, A.~Gurchumelia, Geometry
of the Non-Compact G(2), \emph{J.~Geom.~Phys.}~\textbf{144} (2019), 308-313;
\href{https://arxiv.org/abs/1903.04888}{\emph{arXiv}:1903.04888~[physics.gen-ph]}.

\bibitem[GG2023]{GG2023}M.~Gogberashvili, A.~Gurchumelia, Dirac and
Maxwell systems in split octonions, \emph{J.~Appl.~Math.~Phys.}~\textbf{11}
(2023), 1977--1995; \href{https://arxiv.org/abs/2012.02255}{\emph{arXiv}:2012.02255~[math-ph]}.

\bibitem[GG2024]{GG2024}M.~Gogberashvili, A.~Gurchumelia, Split
octonionic Dirac equation, \emph{Int.~J.~Geom.~Meth.~Mod.~Phys.}\textbf{~21}
(2024), 2450214; \href{https://arxiv.org/abs/2409.13736}{\emph{arXiv}:2409.13736~[physics.gen-ph]}.

\bibitem[Gogb2006a]{Gogb2006a}M.~Gogberashvili, Octonionic Electrodynamics,
\emph{J.~Phys.~A: Math.~Gen.}~\textbf{39} (2006) 7099-7104; \href{https://arxiv.org/abs/hep-th/0512258}{\emph{arXiv}:hep-th/0512258}.

\bibitem[Gogb2006b]{Gogb2006b}M.~Gogberashvili, Octonionic Version
of Dirac Equations, \emph{Int.~J.~Mod.~Phys.~A} \textbf{21} (2006),
3513-3524; \href{https://arxiv.org/pdf/hep-th/0505101}{\emph{arXiv}:hep-th/0505101}.

\bibitem[Gogb2008]{Gogb2008}M.~Gogberashvili, Rotations in the space
of the split octonions, \emph{Adv.~Math.~Phys.}~\textbf{2009} (2009), 483079;
\href{https://arxiv.org/abs/0808.2496}{\emph{arXiv}:0808.2496~[math-ph]}.

\bibitem[HRS2025]{HRS2025}Q.~Huo, G.~Ren, I.~Sabadini, Octonionic
Para-linear Self-Adjoint Operators and Spectral Decomposition, \emph{arXiv}
(2025); \href{https://arxiv.org/abs/2512.04707}{\emph{arXiv}:2512.04707~[math.FA]}.

\bibitem[Jacob1958]{Jacob1958}N.~Jacobson, Composition algebras
and their automorphisms, \emph{Rendiconti del Circolo Matematico di
Palermo} \textbf{7} (1958), 55-80.

\bibitem[KDG2008]{KDG2008}J.~K\"oplinger, V.~Dzhunushaliev, M.~Gogberashvili,
Emergent Time from Non-Associative Quantum Theory, \emph{FQXi essay}
(2008), \href{https://fqxi.org/community/forum/topic/307}{https://fqxi.org/community/forum/topic/307};
\emph{ResearchGate }\href{http://dx.doi.org/10.13140/RG.2.2.10115.99361}{DOI:~10.13140/RG.2.2.10115.99361}.

\bibitem[KHG2025]{KHG2025}J.~K\"oplinger, M.~Habeck, P.~Goyal,
Operational reconstruction of Feynman rules for quantum amplitudes
via composition algebras, \emph{Int.~J.~Theor.~Phys.}~\textbf{65}
(2026), 128; \href{https://arxiv.org/abs/2508.14822}{\emph{arXiv}:2508.14822~[quant-ph]}.

\bibitem[KLL2025]{KLL2025}R.~S.~Krau{\ss}har, A.~Legatiuk, D.~Legatiuk,
Discrete octonionic analysis: a unified approach to the split-octonionic
and classical settings, \emph{arXiv} (2025); \href{https://arxiv.org/abs/2502.02227}{\emph{arXiv}:2502.02227~[math.CV]}.

\bibitem[KLL2025weyl]{KLL2025weyl}R.~S.~Krau{\ss}har, A.~Legatiuk,
D.~Legatiuk, Application of the Weyl Calculus Perspective on Discrete
Octonionic Analysis in Bounded Domains, \emph{Complex Anal.~Oper.~Theory}
\textbf{19} (2025), 26.

\bibitem[KoeplWWW]{KoeplWWW}J.~K\"oplinger, personal web site (publications,
personal versions, talks), \href{http://www.jenskoeplinger.com/P}{http://www.jenskoeplinger.com/P}
(retrieved 1 Nov 2025).

\bibitem[Koepl2006]{Koepl2006}J.~K\"oplinger, Dirac equation on
hyperbolic octonions, \emph{Appl.~Math.~Computation}, \textbf{182}
(2006), 443-446; corrigendum in \cite{Koepl2007c} footnote 2, as
well as \emph{ResearchGate }\href{https://doi.org/10.13140/RG.2.2.22900.41602}{DOI:~10.13140/RG.2.2.22900.41602}.

\bibitem[Koepl2007c]{Koepl2007c}J.~K\"oplinger, Gravity and electromagnetism
on conic sedenions, \emph{Appl.~Math.~Computation}, \textbf{188}
(2007), 948-953.

\bibitem[Koepl2023a]{Koepl2023a}J.~K\"oplinger, Towards autotopies
of normed composition algebras in algebraic Quantum Field Theory,
\emph{ResearchGate} (2023), \href{http://dx.doi.org/10.13140/RG.2.2.19003.39209}{DOI:~10.13140/RG.2.2.19003.39209}.

\bibitem[Koepl2023b]{Koepl2023b}J.~K\"oplinger, Phenomenology from
Dirac equation with Euclidean-Minkowskian \textquotedblleft gravity
phase\textquotedblright , \emph{Int.~J.~Theor.~Phys.}~\textbf{62}
(2023), 35; \href{https://arxiv.org/abs/2302.10748}{\emph{arXiv}:2302.10748~[physics.gen-ph]}.

\bibitem[Lasen2024]{Lasen2024}A.~Lasenby, Some recent results for
SU(3) and Octonions within the Geometric Algebra approach to the fundamental
forces of nature, \emph{Math.~Methods Appl.~Sci\@.} \textbf{47}
(2024), 1471-1491; \href{https://arxiv.org/abs/2202.06733}{\emph{arXiv}:2202.06733~[physics.gen-ph]}.

\bibitem[Lopat2026]{Lopat2026}A.~Lopatin, On polynomial equations
over split-octonions: the arbitrary field case, \emph{Comm.~Math.}
\textbf{34} (2026), Paper No.~12;
\href{https://arxiv.org/abs/2601.07332}{\emph{arXiv}:2601.07332~[math.RA]}.

\bibitem[LR2025]{LR2025}A.~Lopatin, A.~N.~Rybalov, On polynomial
equations over split octonions, \emph{Comm.~Math.} \textbf{33} (2025),
Paper No.~8.

\bibitem[LZ2026]{LZ2024}A.~Lopatin, A.~N.~Zubkov, On linear equations
over split-octonions, \emph{Electronic Journal of Linear Algebra} \textbf{42}
(2026), 569--590; \href{https://arxiv.org/abs/2411.08500}{\emph{arXiv}:2411.08500~[math.RA]}.

\bibitem[MarBon1991]{MarBon1991}S.~Marques-Bonham, The Dirac equation
in a non-Riemannian manifold III: An analysis using the algebra of
quaternions and octonions, \emph{J.~Math.~Phys.}~\textbf{32} (1991),
1383-1394.

\bibitem[MBCM2020]{MBCM2020}S.~Marques-Bonham, B.~C.~Chanyal,
R.~Matzner, Yang-Mills-like field theories built on division quaternion
and octonion algebras, \emph{Eur.~Phys.~J.~Plus} \textbf{135} (2020),
608.

\bibitem[MD1999]{MD1999}C.~A.~Manogue, T.~Dray, Dimensional Reduction,
\emph{Mod.~Phys.~Lett.~}\textbf{A14} (1999) 99-103; \href{https://arxiv.org/abs/hep-th/9807044}{\emph{arXiv}:hep-th/9807044}.

\bibitem[MDW2022]{MDW2022}C.~A.~Manogue, T.~Dray, R.~A.~Wilson,
Octions: An $E_{8}$ description of the Standard Model, \emph{J.~Math.~Phys.}~\textbf{63}
(2022), 081703; \href{https://arxiv.org/abs/2204.05310}{\emph{arXiv}:2204.05310~[hep-ph]}.

\bibitem[Muses1980]{Muses1980}C.~Mus\`es, Hypernumbers and quantum
field theory with a summary of physically applicable hypernumber arithmetics
and their geometries, \textit{Appl.~Math.~Comput.}~\textbf{6} (1980),
63-94.

\bibitem[Okubo1995]{Okubo1995}S.~Okubo, Introduction to Octonion
and Other Non-Associative Algebras in Physics (1995), Cambridge Univ.~Press.

\bibitem[Penney1968]{Penney1968}R.~Penney, Octonions and the Dirac
equation, \emph{Am.~J.~Phys.}~\textbf{36} (1968), 871-873.

\bibitem[Penney1971]{Penney1971}R.~Penney, Octonions and isospin,
\emph{Nuovo Cimento B} \textbf{3} (1971), 95-113.

\bibitem[SM1996]{SM1996}J.~Schray, C.~Manogue, Octonionic representations
of Clifford algebras and triality, \emph{Found.~Phys.~}\textbf{26}
(1996) 17-70; \href{https://arxiv.org/abs/hep-th/9407179}{\emph{arXiv}:hep-th/9407179}.

\bibitem[Sudb1984]{Sudb1984}A.~Sudbery, Division algebras, (pseudo)orthogonal
groups and spinors, \emph{J.~Phys.~A} \textbf{17}
(1984), 939-955.

\bibitem[TKD2012]{TKD2012}M.~Tan\i\c{s}l\i, M.~E.~Kansu, S.~Demir,
A new approach to Lorentz invariance in electromagnetism with hyperbolic
octonions, \emph{Eur.~Phys.~J.~Plus} \textbf{127} (2012), 69.

\bibitem[TKD2014]{TKD2014}M.~Tan\i\c{s}l\i, M.~E.~Kansu, S.~Demir,
Reformulation of electromagnetic and gravito-electromagnetic equations
for Lorentz system with octonion algebra, \emph{Gen.~Relativ.~Gravit.~}\textbf{46}
(2014), 1739.

\bibitem[VQ2022]{VQ2022}J.~C.~V\'elez Qui\~nones, A unified field
theory from a complexified quaternion-octonion Dirac equation, \emph{arXiv}
(2022); \href{https://arxiv.org/abs/2205.06657}{\emph{arXiv}:2205.06657~[physics.gen-ph]}.

\bibitem[Weng2009]{Weng2009}Z.~Weng, Field equations of electromagnetic
and gravitational fields, \emph{arXiv} (2009); \href{https://arxiv.org/abs/0709.2486}{\emph{arXiv}:0709.2486~[physics.gen-ph]}.

\end{thebibliography}
\end{document}